\documentclass[12pt]{article}

\usepackage[margin=1in]{geometry}
\usepackage{graphics,url}
\usepackage{pstricks,pst-node,pst-tree}
\usepackage{amsmath,url}

\usepackage[numbers, sort&compress]{natbib}
\newtheorem{algorithm}{Algorithm}

\title{A regression tree approach to identifying subgroups\\
  with differential treatment effects}

\author{Wei-Yin Loh\\ \texttt{loh@stat.wisc.edu} \\
  Department of Statistics, University of Wisconsin \\
  Madison, WI 53706, U.S.A.  \and
  Xu He\\ \texttt{hexu@amss.ac.cn} \\
  Academy of Mathematics and Systems Science\\ Chinese Academy of
  Sciences, Beijing, China \and
  Michael Man\\ \texttt{man\_michael@lilly.com} \\
  Eli Lilly and Company\\
  Indianapolis, IN 46285, U.S.A.  \and
}

\date{}

\begin{document}


\maketitle

\begin{abstract}
  In the fight against hard-to-treat diseases such as cancer, it is
  often difficult to discover new treatments that benefit all
  subjects.  For regulatory agency approval, it is more practical to
  identify subgroups of subjects for whom the treatment has an
  enhanced effect.  Regression trees are natural for this task because
  they partition the data space. We briefly review existing regression
  tree algorithms. Then we introduce three new ones that are
  practically free of selection bias and are applicable to two or more
  treatments, censored response variables, and missing values in the
  predictor variables.  The algorithms extend the GUIDE approach by
  using three key ideas: (i) treatment as a linear predictor, (ii)
  chi-squared tests to detect residual patterns and lack of fit, and
  (iii) proportional hazards modeling via Poisson regression.
  Importance scores with thresholds for identifying influential
  variables are obtained as by-products. And a bootstrap technique is
  used to construct confidence intervals for the treatment effects in
  each node.  Real and simulated data are used to compare the methods.


\end{abstract}

\textbf{Key words:} Missing values; proportional hazards; selection
bias.

\section{Introduction}
\label{sec:intro}

For many diseases, such as cancer, it is often difficult to find a
treatment that benefits all patients.  Current thinking in drug
development is to find a subject subgroup, defined by individual
characteristics, that shows a large treatment effect. Conversely, if a
treatment is costly or has potential negative side effects, there is
interest to look for subgroups for which it treatment is ineffective.
This problem of searching for subgroups with differential treatment
effects is known as \emph{subgroup identification}
\cite{ciampi95,FTR11,Lipkovich11}.

To fix ideas, suppose for the moment that the response variable $Y$ is
uncensored and the treatment variable $Z$ takes values $l = 1, 2,
\ldots, L$. Let $\mathbf{X}$ denote a vector of covariates. Given a
subgroup $S$ defined in terms of $\mathbf{X}$, let $R(S) = \max_{i,j}
|E(Y| Z=i, S) - E(Y| Z=j, S)|$ denote the effect size of $S$. The goal
is to find the maximal subgroup with the largest value of $R(S)$,
where the size of $S$ is measured in terms of its probability of
occurrence $P(S)$. If $Y$ is subject to censoring, we replace the mean
of $Y$ by the log-hazard rate so that $R(S)$ is the largest absolute
log-hazard ratio between any two treatments.

Consider, for example, data from a randomized trial of the German
Breast Cancer Study Group \cite{schmoor96,ipred} with 686 subjects
where the response is recurrence-free survival time in days.  The
trial was designed as a $2 \times 2$ factorial comparing 3 vs.\ 6
cycles of chemotherapy and presence vs.\ absence of hormone therapy
(Tamoxifen).  The breast cancer data contain, however, no information
on the number of cycles of chemotherapy, presumably because it was
previously found not significant \cite{schumacher94}.  Median
follow-up time was nearly 5 years and 387 subjects did not experience
a recurrence of the disease during the trial (54 percent
censoring). The variables are hormone therapy (\texttt{horTh}: yes,
no), age (21--80 years), tumor size (\texttt{tsize}: 3--120 mm),
number of positive lymph nodes (\texttt{pnodes}: 1--51), progesterone
receptor status (\texttt{progrec}: 0--2380 fmol), estrogen receptor
status (\texttt{estrec}: 0--1144 fmol), menopausal status
(\texttt{menostat}: pre, post), and tumor grade (\texttt{tgrade}: 1,
2, 3).  A standard proportional hazards regression model shows that
hormone therapy has a significant positive effect on survival, with
and without adjusting for the covariates \cite{sauerbrei99,HSAUR}.
Since hormone therapy has side effects and incurs extra cost, it is
useful to find a subgroup where the treatment has little effect.

Parametric and semi-parametric models such as the proportional hazards
model do not easily lend themselves to this problem.  Besides, if
there are more variables than observations, such as in genetic data,
these models cannot be used without prior variable selection.
Regression tree models are better alternatives, because they are
nonparametric, naturally define subgroups, scale with the complexity
of the data, and are not limited by the number of predictor variables.

\begin{figure}
  \centering
  \psset{linecolor=black,tnsep=2pt,tnheight=0cm,treesep=2cm,levelsep=40pt,radius=10pt}
  \mbox{} \hfill
  \pstree{\TC~[tnpos=l]{pnodes $\leq$ 3}~[tnpos=a]{\Large Gs}}{
    \TC[fillcolor=cyan,fillstyle=solid]~{\makebox[0pt][c]{(0.33, 1.02)}}
    ~[tnpos=l]{370}
    \TC[fillcolor=cyan,fillstyle=solid]~{\makebox[0pt][c]{(0.46, 1.06)}}
    ~[tnpos=r]{302}
  }
  \hfill \mbox{} \hfill
  \pstree{\TC~[tnpos=l]{progrec $\leq$ 21}~[tnpos=a]{\Large Gi}}{
    \TC[fillcolor=cyan,fillstyle=solid]
    ~{\makebox[0pt][c]{(0.55, 1.42)}}~[tnpos=l]{274}
    \TC[fillcolor=cyan,fillstyle=solid]~{\makebox[0pt][c]{(0.29, 0.94)}}
    ~[tnpos=r]{398}
  }
  \hfill \mbox{}
  \\ \vspace{1em}
  \resizebox{0.8\textwidth}{!}{\rotatebox{-90}{\includegraphics*[33,19][577,777]{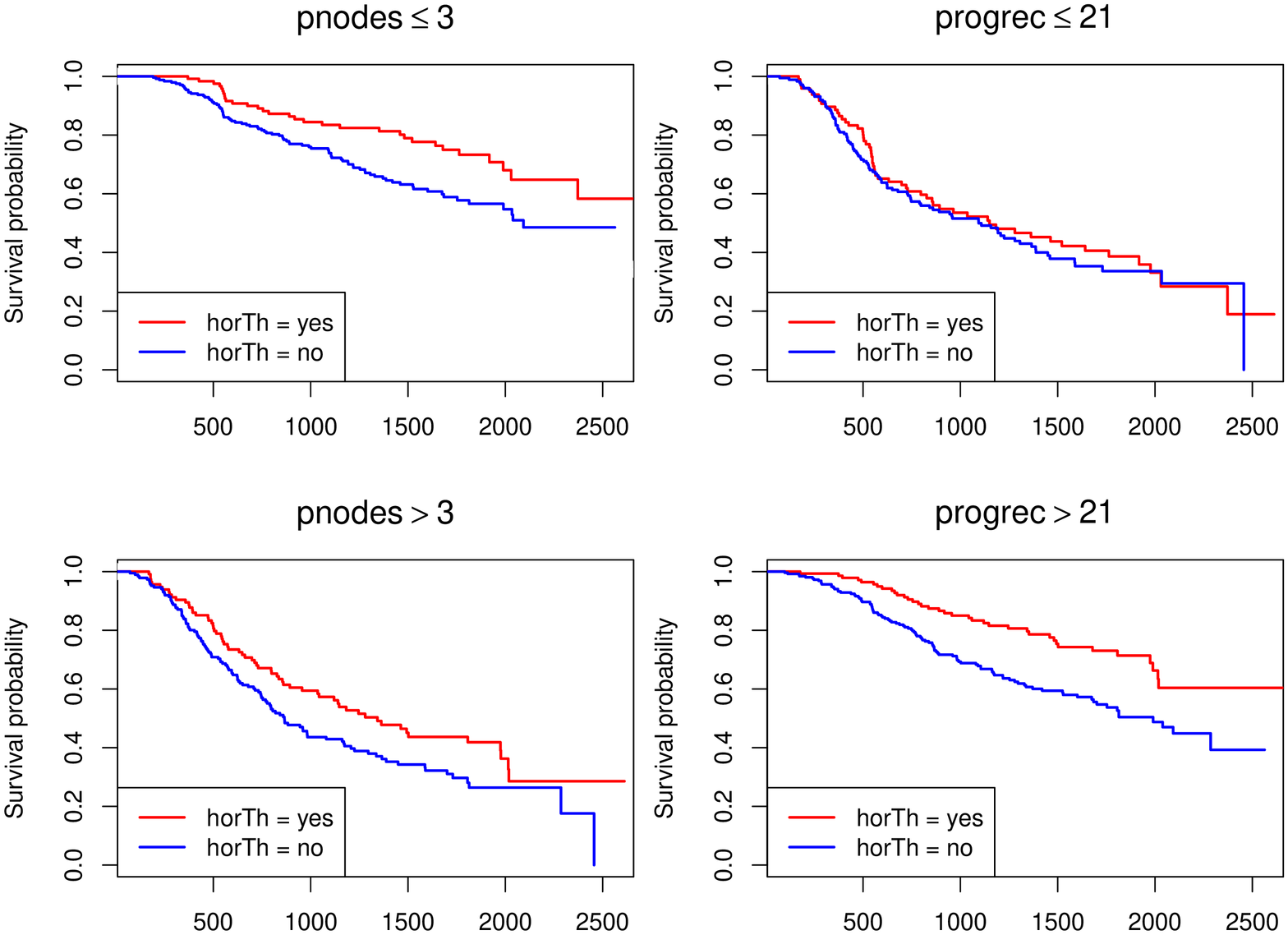}}}
  \caption{Gs (left) and Gi (right) tree models and Kaplan-Meier
    curves for breast cancer data. At each intermediate node, an
    observation goes to the left child node if and only if the
    displayed condition is satisfied.  Sample sizes are beside
    terminal nodes; 95\% bootstrap confidence intervals of relative
    risks for therapy versus no therapy are below nodes.}
  \label{fig:breast:gsgi}
\end{figure}

Following medical practice \cite{Italiano11}, we call a variable
\emph{prognostic} if it provides information about the response
distribution of an untreated subject. That is, it has marginal effects
on the response but does not interact with treatment.  Examples are
age, family history of disease, and prior therapy.  A variable is
\emph{predictive} if it defines subgroups of subjects who are more
likely to respond to a given treatment. That is, it has interaction
effects with the treatment variable. Figure~\ref{fig:breast:gsgi}
shows two regression tree models and the Kaplan-Meier curves in the
terminal nodes of the trees. In the Gs model on the left, variable
\texttt{pnodes} is prognostic: recurrence probability is reduced if
\texttt{pnodes} $>$ 3, with and without treatment.  In the Gi model on
the right, variable \texttt{progrec} is predictive: hormone therapy
has little effect if \texttt{progrec} $\leq$ 21 and an enhanced effect
otherwise.

The main goal of this article is to introduce the algorithms that
yield the models in Figure~\ref{fig:breast:gsgi} and to compare them
against existing solutions, which are briefly reviewed in
Sec.~\ref{sec:previous}.  Sec.~\ref{sec:uncensored} presents the new
algorithms for uncensored response variables.  Sec.~\ref{sec:bias-acc}
compares the selection bias and accuracy of the new and old methods
and Sec.~\ref{sec:boot} proposes a bootstrap technique for computing
confidence intervals of treatment and other effects in the
subgroups. Sec.~\ref{sec:censored} extends the algorithms to censored
survival data and Sec.~\ref{sec:imp} obtains importance scores for
ranking the variables and and thresholds for identifying the
unimportant ones. Sec.~\ref{sec:gene} gives an application to a
retrospective candidate gene study where there are large numbers of
missing values and Sec.~\ref{sec:conc} concludes the article with some
closing remarks.

\section{Previous work}
\label{sec:previous}
Let $y_i$ and $\mathbf{x}_i$ denote the survival time and covariate
vector of subject~$i$.  Let $s_i$ be an independent observation from
some censoring distribution and let $\delta_i = I(y_i < s_i)$ be the
event indicator.  The observed data vector of subject $i$ is $(y_i,
\delta_i, \mathbf{x}_i)$, where $y_i = \min(y_i, s_i)$, $i=1,2,\ldots,
n$.  Let $\lambda(y, \mathbf{x})$ denote the hazard function at
$\mathbf{x} = (x_1, x_2, \ldots, x_M)$. The proportional hazards model
specifies that $\lambda(y, \mathbf{x}) = \lambda_0(y) \exp(\eta)$,
where $\lambda_0(y)$ is a baseline hazard and $\eta =
\mathbf{\beta}'\mathbf{x}$ is a linear function of the covariates.



Assuming that the treatment has two levels (denoted by $z=0, 1$), one
approach \cite{NCASB05} splits each node $t$ into left and right child
nodes $t_L$ and $t_R$ to maximize the Cox partial likelihood ratio
statistic for testing $H_0 : \lambda(y,\mathbf{x}) = \lambda_{0,t}(y)
\exp\{\beta_0 z I(\mathbf{x} \in t)\}$ against $H_1 :
\lambda(y,\mathbf{x}) = \lambda_{0,t}(y) \exp\{\beta_1 z I(\mathbf{x}
\in t_L) + \beta_2 z I(\mathbf{x} \in t_R) \}$.  A related approach,
called \emph{interaction trees} (IT) \cite{SZYFY08}, chooses the split
that minimizes the p-value from testing $H_0: \beta_3 = 0$ in the
model $\lambda(y,\mathbf{x}) = \lambda_{0,t}(y) \exp\{\beta_1 z +
\beta_2 I(\mathbf{x} \in t_L) + \beta_3 z I(\mathbf{x} \in t_L)\}$.
If there is no censoring, the model is $E(y) = \beta_0 + \beta_1 z +
\beta_2 I(\mathbf{x} \in t_L) + \beta_3 z I(\mathbf{x} \in t_L)$.
Both methods employ the greedy search paradigm of evaluating all
splits $t_L = \{x_j \in S\}$ and $t_R = \{x_j \not\in S\}$ on every
$x_j$ and every $S$, where $S$ is a half line if $x_j$ is ordinal and
is a subset of values if $x_j$ is categorical. As a result, they are
computationally expensive and biased toward selecting variables that
allow more splits.  Further, because $\lambda_{0,t}(y)$ is a function
of $t$ and hence of $\mathbf{x}$, the tree models do not have
proportional hazards and regression coefficients in different nodes
cannot be compared.

The \emph{virtual twins} (VT) method \cite{FTR11} is restricted to
binary variables $Y = 0, 1$. It first generates a random forest
\cite{rf01} model to estimate the treatment effect $\tau =
P(Y=1\,|\,Z=1) - P(Y=1\,|\,Z=0)$ of each subject, using as split
variables $Z, X_1, \ldots, X_M, ZX_1, \ldots, ZX_M, (1-Z)X_1, \ldots,
(1-Z)X_M$, with categorical variables converted first to dummy 0-1
variables.  Then it uses RPART \cite{rpart} to construct a
classification or regression tree model to predict $\tau$ for each
subject and to obtain the subgroups. If a classification tree is used,
the two classes are defined by the estimated $\tau$ being greater or
less than a pre-specified constant; if a regression tree is used, the
subgroups are the terminal nodes with estimated $\tau$ greater than a
pre-specified constant. Although the basic concept is independent of
random forest and RPART, their use gives VT all their weaknesses, such
as variable selection bias and (with random forest) lack of a
preferred way to deal with missing values. Further, VT is limited to
binary $Y$ and $Z$.

The \emph{subgroup identification based on differential effect search}
(SIDES) method \cite{Lipkovich11} finds multiple alternative
subgroups by identifying the best five (default) splits of each node
that yield the most improvement in a desired criterion, such as the
p-values of the differential treatment effects between the two child
nodes, the treatment effect size in at least one child node, or the
difference in efficacy and safety between the two child nodes.  For
each split, the procedure is repeated on the child node with the
larger improvement.  Heuristic and resampling-based adjustments are
applied to the p-values to control for multiplicity of splits and
correlations among the p-values.  The method appears to be most useful
for generating candidate subgroups with large differential effects,
but because only variables that have not been previously chosen are
considered for splitting each node, the method may not be effective if
the real subgroups are defined in terms of interval sets of the form
$\{a_j < X_j \leq b_j\}$.  The current implementation is limited to
treatments with two levels.

Most methods can control the minimum node sample size so that the
subgroups have sufficient numbers of observations.  The
\emph{qualitative interaction tree} (QUINT) method \cite{dusseldorp13}
deals with this directly by optimizing a weighted sum of a measure of
effect size and a measure of subgroup size.  It looks for
``qualitative interactions,'' where one treatment performs better than
another in one subgroup and worse in another subgroup. Similar to the
above methods, QUINT finds the subgroups by searching over all
possible splits on all predictor variables. Its current implementation
is limited to ordinal $X_i$, uncensored $Y$, and binary $Z$.

\section{Uncensored data}
\label{sec:uncensored}
By evaluating all possible splits on all variables to optimize an
objective function, each method (except possibly for SIDES) is biased
toward selecting variables that allow more splits.  This is due to an
ordinal variable with $k$ unique values yielding $k-1$ splits and a
categorical variable with the same number of unique values yielding
$2^{k-1}-1$ splits. As a result, a variable that allows many splits
has a greater chance to be selected than one with few splits. Besides
increasing the chance of spurious splits, the bias can undermine the
credibility of the results.  SIDES tries to control the bias with
Bonferroni-type adjustments, but this can lead to over correction, as
in the CHAID \cite{Kass80} classification tree algorithm, which is
biased toward selecting variables with few splits. 

The GUIDE algorithm \cite{guide02,guide09} overcomes this problem by
using a two-step approach to split selection: first find the split
variable and then search for the best split on the selected
variable. The first step yields substantial computational savings,
because there is no need to find the best splits on the all the other
variables.  It also eliminates selection bias, at least in principle,
by using chi-squared tests to select the split variable.  QUEST
\cite{quest}, CRUISE \cite{cruise}, and CTREE \cite{party} are other
algorithms that employ significance tests for variable selection.  In
this section we extend GUIDE to subgroup identification for the case
where $Y$ is not censored.

\subsection{Gc: classification tree approach}
\label{sec:gz}
This method requires that $Y$ and $Z$ are binary, taking values, 0,
and 1, say. Then a classification tree may be used to find subgroups
by defining the class variable as $V = Y+Z \bmod{2}$:
\[ V=\left\{ \begin{array}{ll}
    0, & \mbox{ if $\{Y=1$ and $Z=1\}$ or $\{Y=0$ and $Z=0\}$,} \\
    1, & \mbox{ if $\{Y=0$ and $Z=1\}$ or $\{Y=1$ and $Z=0\}$.} 
  \end{array}\right.
\]
This is motivated by the observation that the subjects for which $V =
0$ respond differentially to treatment and those for which $V=1$ do
not.  Thus a classification tree constructed with $V$ as the response
variable will likely identify subgroups with differential treatment
effects. Although any classification tree algorithm may be used, we
use GUIDE \cite{guide09} here because it does not have selection bias,
and call it the Gc method (``c'' for classification).

\subsection{Gs and Gi: regression tree approach}
\label{sec:g14}

\begin{figure}[htbp]
  \begin{center}    
    \resizebox{0.75\textwidth}{!}{\includegraphics*[24,397][576,736]{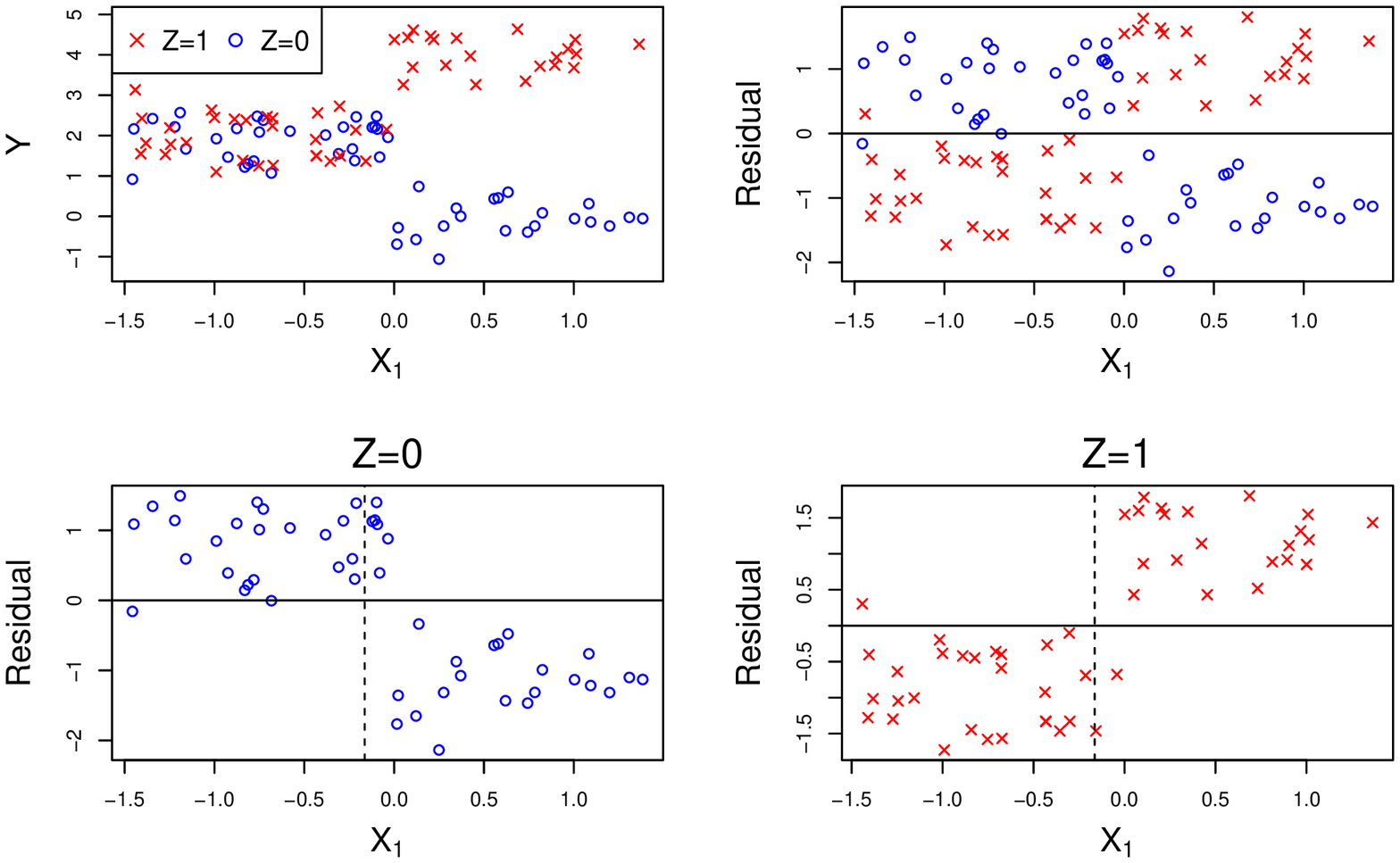}}\\
    \vspace{1em}
    \mbox{} \hfill \mbox{} \hfill \mbox{} \hfill
    \begin{minipage}{2.5in}      
      \begin{tabular}{l|rr}
        $Z = 0$ & $X_1 \leq \bar{x}_1$ & $X_1 > \bar{x}_1$ \\ \hline
        resid $>$ 0 & 21 & 6 \\
        resid $\leq$ 0 & 2 & 21 \\ \hline        
        \multicolumn{3}{c}{$\chi^2 = 21.2$, $p = 4 \times 10^{-6}$}
      \end{tabular}
    \end{minipage}
    \hfill
    \begin{minipage}{2.5in}      
      \begin{tabular}{l|rr}
        $Z = 1$ & $X_1 \leq \bar{x}_1$ & $X_1 > \bar{x}_1$ \\ \hline
        resid $>$ 0 & 1 & 21 \\
        resid $\leq$ 0 & 26 & 2 \\ \hline        
        \multicolumn{3}{c}{$\chi^2 = 35.2$, $p = 3 \times 10^{-9}$}
      \end{tabular}
    \end{minipage}
    \hfill \mbox{}
  \end{center}
  \caption{Plots of $Y$ and residuals vs.\ $X_1$ after fitting $EY =
    \beta_0 + \beta_1Z$ to data from model~(\ref{eq:purepred});
    vertical dashed lines indicate sample mean of $X_1$}
  \label{fig:regex:pred}
\end{figure}

The GUIDE linear regression tree algorithm \cite{guide02} provides an
alternative approach that permits $Y$ and $Z$ to take more than two
values. At each node, we fit a model linear in $Z$ and select the
variable to split it by examining the residual patterns for each level
of $Z$.  Consider, for example, data $(X_1, X_2, \ldots, Y, Z)$
generated from the model
\begin{equation}
  \label{eq:purepred}
  Y = 1.9 + 0.2I(Z=1) - 1.8I(X_1 >0) + 3.6I(X_1 > 0, Z=1) + \varepsilon
\end{equation}
where $Z = 0, 1$ and $\varepsilon$ is independent normal.  Because the
true subgroup is $X_1 > 0$, we should split the data using $X_1$.
Figure~\ref{fig:regex:pred} shows how this conclusion can be reached
from the data alone. The top row plots $Y$ and the residuals vs.\
$X_1$, where the residuals are from fitting the model $EY = \beta_0 +
\beta_1 Z$. The middle row plots residuals vs.\ $X_1$ for each $Z$
level. The distinct nonrandom patterns in the latter plots are
indicators that $X_1$ has an interaction with $Z$. No other variables
can be expected to show such strong patterns. We measure the strength
of the interaction by forming a contingency table with the residual
signs as rows and grouped values of $X_1$ (obtained by dividing its
values at the sample mean) as shown in the bottom row of the figure,
computing a chi-squared statistic for each level of $Z$, and summing
them.  Repeating this procedure for each $X_i$, we rank the variables
and select the one with the largest summed chi-squared to split the
data.  We call this the Gs method (``s'' for sum).

Contingency table tests are convenient because they are quick to
compute, can detect a large variety of patterns, and are applicable to
categorical $X$ variables, where we use their values for the columns.
Because the latter changes the degrees of freedom (df) of the
chi-squared statistics, we need to adjust for differences in df before
summing them. We do this by following GUIDE which uses a double
application of the Wilson-Hilferty approximation \cite{wh31}
to convert each contingency table chi-squared statistic to a 1-df
chi-squared quantile.  Specifically, let $x$ and $y$ be chi-squared
quantiles with $\nu$ df and $\mu$, respectively, degrees of
freedom. Then (see \cite{guide09})
\begin{equation}
  \label{eq:wh}
  y \approx \mu \left[ 1 - 2/(9\mu) + \sqrt{\nu/\mu} 
    \{ (x/\nu)^{1/3} - 1 + 2/(9\nu)\} \right]^3.  
\end{equation}
After a variable is selected, a search is carried out for the best
split on the variable that minimizes the sum of squared residuals in
the two child nodes and the process is applied recursively to each
node. The detailed algorithm, including handling of missing values, is
given below.

\begin{algorithm} Gs split selection. \label{alg:Gs}
\begin{enumerate}
\item Fit the least squares model $EY = \beta_0 + \sum_{z=1}^L \beta_z
  I(Z = z)$ to the data in the node and compute the residuals. Let
  $S_z$ denote the set of observations with $Z = z$ in the node.
\item For each $X$ and $z=1,2,\ldots, L$.
  \begin{enumerate}
  \item \label{step:cont} Form a contingency table from the data in
    $S_z$ using the signs (positive vs.\ non-positive) of the
    residuals as columns and the (grouped) $X$ values as rows. If $X$
    is ordinal, divide its values into two groups at the mean.
    Otherwise, if $X$ is categorical, let its values define the
    groups. If there are missing values, add an additional ``missing
    value'' group.
  \item Compute the chi-squared statistic $W_z$ for testing
    independence, and let $\nu_z$ denote its degrees of freedom.  Use
    (\ref{eq:wh}) to convert $W_z$ to the 1-df chi-squared quantile
    \[ r_z(X) = \max \left(0, \left[ 7/9 + \sqrt{\nu_z} \left\{
          \left( W_z/\nu_z \right)^{1/3} - 1 +
          2/(9\nu_z) \right\} \right]^3 \right). \]
  \end{enumerate}
\item Treating $\sum_{z=1}^L r_z(X)$ as a chi-squared variable with
  $L$ df, use (\ref{eq:wh}) a second time to convert it to a 1-df
  chi-squared quantile
  \[ q(X) = \max \Big(0, \Big[ 7/9 + \sqrt{L} \Big\{ \big( L^{-1}
  \sum\nolimits_{z} r_z(X) \big)^{1/3} - 1 + 2/(9L) \Big\} \Big]^3
  \Big). \]
\item Let $X^*$ be the variable with the largest value of $q(X)$. \label{step:spltpt}
  \begin{enumerate}
  \item If $X^*$ is ordinal, let $A$ denote the event that $X^*$ is
    missing (if any) and $\bar{A}$ its complement.  Then search
    through the values of $c$ for the split $A \cap \{X^* \leq c\}$ or
    $\bar{A} \cap \{X^* \leq c\}$ that minimizes the sum of the
    squared residuals fitted to the two child nodes produced by the
    split.
  \item If $X^*$ is categorical, let $g$ denote its number of
    categories (including the missing category, if any). If $g < 10$,
    search over all $(2^{g-1}-1)$ splits of the form $X^* \in S$ to
    find the one that minimizes the sum of squared residuals in the
    two child nodes.  If $g \geq 10$, limit the search to $(g-1)$
    splits as follows.
    \begin{enumerate}
    \item Label an observation as belonging to class~1 if it has a
      positive residual and as class~2 otherwise.
    \item Order the $X^*$ values by their proportions of class~1
      subjects in the node.
    \item Select the split along the ordered $X^*$ values that yields
      the greatest reduction in the sum of Gini indices. (This mimics
      a technique in \cite[p.~101]{cart} for piecewise constant
      least-squares regression.)
    \end{enumerate}
  \end{enumerate}
  If there are no missing $X^*$ values in the training data, future
  cases missing $X^*$ are sent to the more populous child node.
\end{enumerate}
\end{algorithm}

\begin{figure}
  \centering
  \resizebox{0.75\textwidth}{!}{\includegraphics*[23,396][577,737]{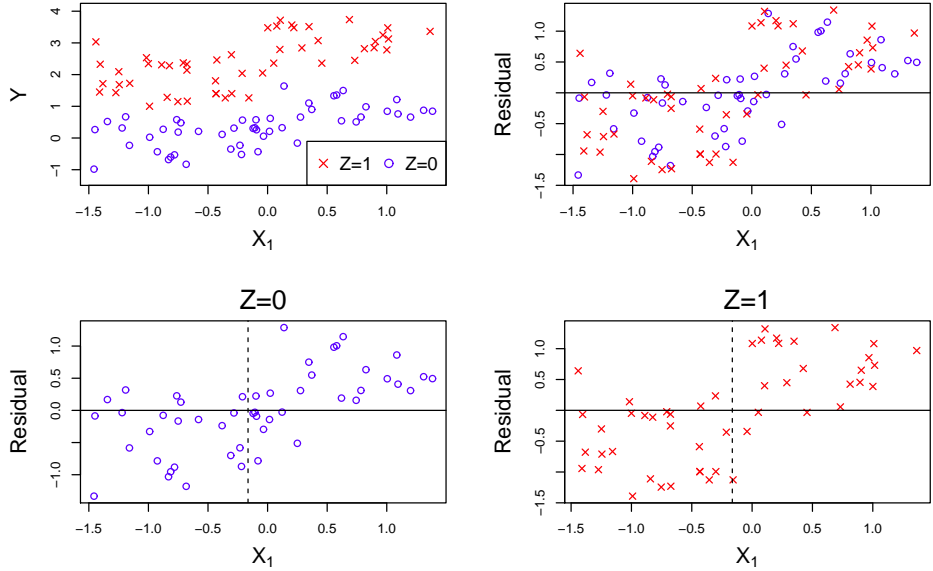}}
  \caption{Plots of data and residuals from the
    model~(\ref{eq:pureprog}) where $X_1$ is prognostic}
  \label{fig:regex:prog}
\end{figure}

Because Gs (as well as MOB) is sensitive to both prognostic and
predictive variables, it may be ineffective if only subgroups defined
by predictive variables are desired.  To see this, suppose now that
the data $(X_1, X_2, \ldots, Y, Z)$ are generated from the true model
\begin{equation}
  \label{eq:pureprog}
  Y = 2 I(Z=1) + I(X_1 > 0) + \varepsilon
\end{equation}
with $\varepsilon$ independent normal.  The simulated data plots in
Figure~\ref{fig:regex:prog} show that Gs will choose $X_1$ with high
probability even though it is prognostic but not predictive.  IT
overcomes this by adding the interaction $I(Z=1)I(X > c)$ to the
fitted model and testing for its significance, but this approach
requires searching over the values of $c$, which produces selection
bias and may be impractical if $Z$ takes more than two levels. To get
around these problems, we instead test for lack of fit of the model
\begin{equation}
  \label{eq:additive}
  EY = \beta_0 + \sum_k \beta_k I(Z = k) + \sum_j \gamma_j I(H=j)
\end{equation}
where $H=X$ if it is categorical and is the indicator function $I(X
\leq \bar{x})$ with $\bar{x}$ being the sample mean of $X$ at the node
otherwise. Then we select the most significant $X$ to split the data.
Turning an ordinal $X$ into a binary variable may lead to loss of
power, but this is compensated by increased sensitivity to
interactions of all kinds, including those that cannot be represented
by cross-products of indicators.  We call this the Gi method (``i''
for interaction). The procedure, including handling of missing values,
is given next.


\begin{algorithm} Gi split selection.
  \label{alg:Gi}
  \begin{enumerate}
  \item For each $X$ variable at each node:
    \begin{enumerate}
    \item If $X$ is ordinal, divide its values into two groups at its
      mean. If $X$ is categorical, let its values define the
      groups. Add a group for missing values if there are any. Let $H$
      denote the factor variable created from the groups.
    \item Carry out a lack-of-fit test of the
      model~(\ref{eq:additive}) on the data in the node and convert
      its p-value to a 1 df chi-squared statistic $q(X)$.
    \end{enumerate}
  \item Let $X^*$ be the variable with the largest value of $q(X)$ and
    use the procedure in Algorithm~\ref{alg:Gs} step~\ref{step:spltpt}
    to find the split on $X^*$ that minimizes the sum of squared
    residuals of the model $EY = \eta + \sum_k \beta_k I(Z = k)$
    fitted to the child nodes.
  \end{enumerate}
\end{algorithm}

\section{Bias and accuracy} \label{sec:01:compare} 
\label{sec:bias-acc}

\subsection{Selection bias}
\label{sec:01:bias}

It is obviously important for a tree model not to have selection bias
if it is used for subgroup identification.  At the minimum, this
requires that if all the variables are independent of $Y$, each $X_i$
has the same probability of being selected to split each node. We
carried out a simulation experiment to compare the methods on this
score. The experiment employed two predictors, $X_1$ and $X_2$, and
Bernoulli response and treatment variables $Y$ and $Z$ each with
success probability 0.50. All variables are mutually independent.  The
distributions of $X_1$ and $X_2$ ranged from standard normal, uniform
on the integers 1, 2, 3, 4, and equi-probable categorical with 3 and 7
levels, as shown in Table~\ref{tab:bias:d}.

\begin{figure}
  \centering
  \resizebox{0.8\textwidth}{!}{\rotatebox{-90}{\includegraphics*[43,37][585,751]{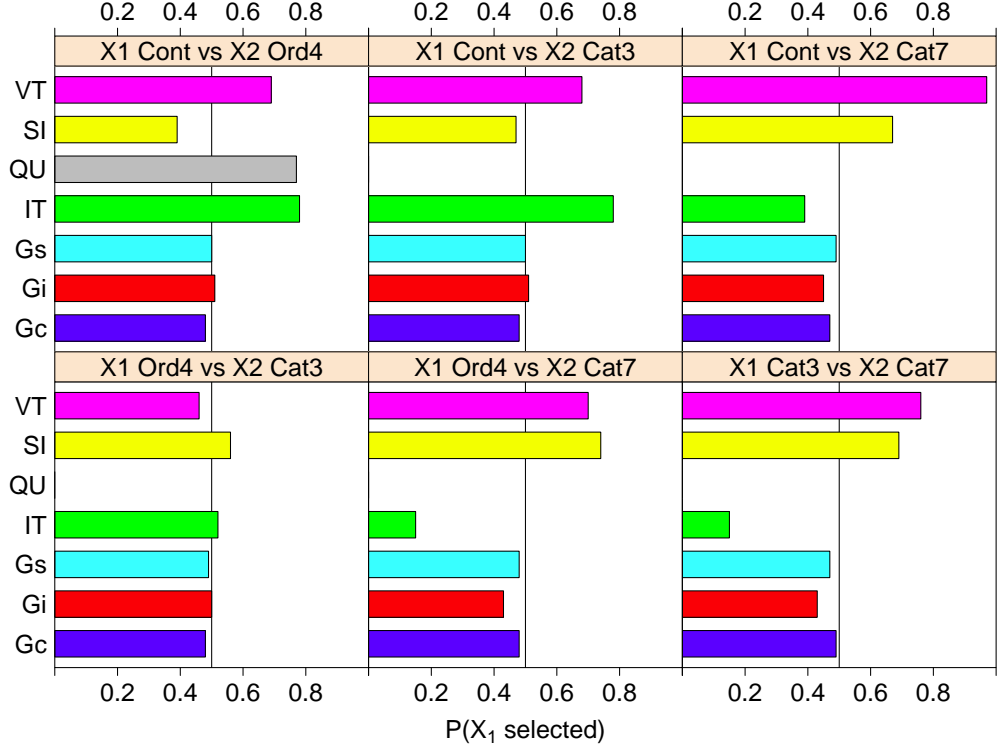}}}  
  \caption{Simulated probabilities that $X_1$ is selected to split the
    root node; simulation standard errors less than 0.01. A method is
    unbiased if it selects $X_1$ with probability 0.50. SI and QU
    refer to SIDES and QUINT, respectively. The latter is not
    applicable to categorical variables.}
  \label{fig:bias:01}
\end{figure}

Based on a sample size of 100 for each iteration,
Figure~\ref{fig:bias:01} shows the frequency that each method selects
$X_1$ to split the root node over 2500 simulation iterations.
Simulation standard errors are less than 0.01.  An unbiased method
should select $X_1$ or $X_2$ with equal probability regardless of
their distributions.  The results show that IT, QUINT, SIDES and VT
have substantial selection biases (QUINT is limited to ordinal $X_i$).
IT and QUINT are biased towards selecting the variable that has more
splits while SIDES and VT are opposite.  In contrast, the selection
frequencies of Gs and Gc are all within three simulation standard
errors.  The frequencies of Gi are also within three standard errors,
except when $X_2$ is categorical with 7 levels where it has a slightly
higher chance to be selected.


\subsection{Accuracy} \label{sec:01:power} 

\begin{figure}
  \centering
  \resizebox{\textwidth}{!}{\rotatebox{-90}{\includegraphics*[23,24][269,754]{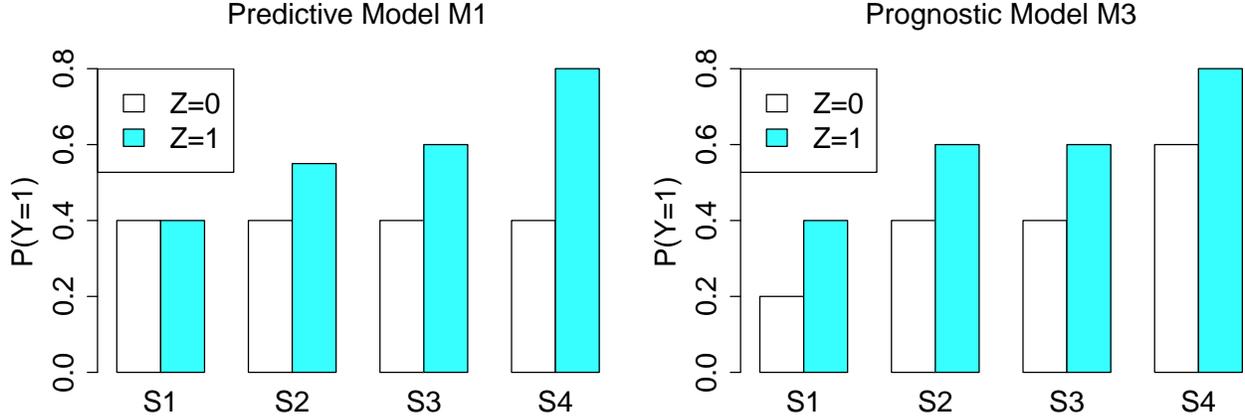}}}  
  \caption{Models~M1 and M3. Sets S1 = $\{X_1=0,X_2=0\}$, S2 =
    $\{X_1=0,X_2> 0\}$, S3 = $\{X_1> 0,X_2=0\}$, and S4 =
    $\{X_1> 0,X_2> 0\}$.}
  \label{fig:models:m1&3}
\end{figure}

\begin{figure}
  \centering
  \resizebox{\textwidth}{!}{\rotatebox{-90}{\includegraphics*[18,23][278,762]{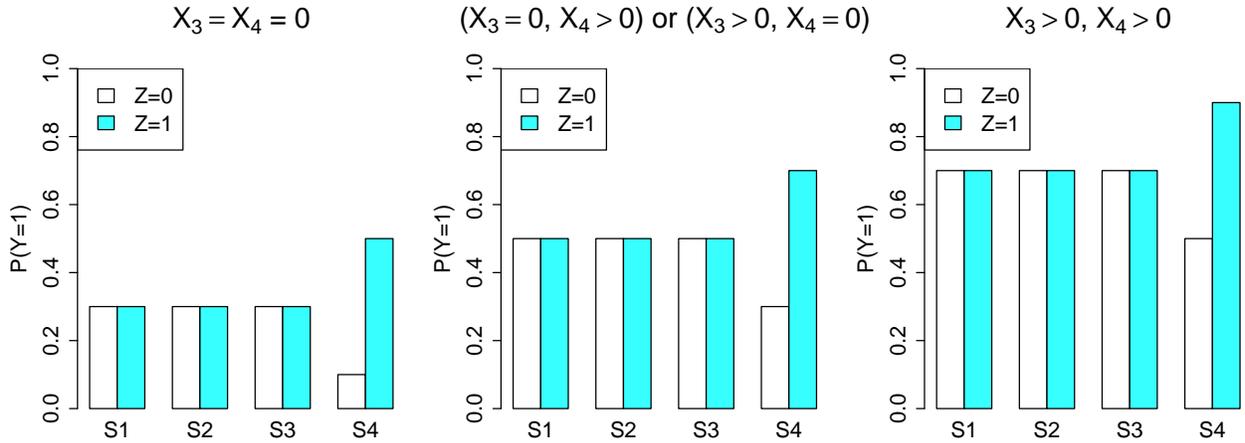}}}  
  \caption{Model M2, with predictive $X_1$ and $X_2$ and prognostic
    $X_3$ and $X_4$. Sets S1 = $\{X_1=0,X_2=0\}$, S2 =
    $\{X_1=0,X_2> 0\}$, S3 = $\{X_1> 0,X_2=0\}$, and S4 =
    $\{X_1> 0,X_2> 0\}$.}
  \label{fig:models:m2}
\end{figure}

We use three simulation models to compare the methods in terms of
their accuracy in selecting the correct variables and the correct
subgroups.  Each model employs a binary treatment variable $Z$ with
$P(Z=0) = P(Z=1)$ and 100 variables $\mathbf{X} = (X_1, X_2, \ldots,
X_{100})$, all mutually independent. Each $X_i$ takes categorical
values 0, 1, or 2 (simulating genetic markers with genotypes AA, Aa,
and aa), with $X_1$ and $X_2$ having identical marginal distribution
$P(X_1=0)=0.4$, $P(X_1=1)=0.465$ and $P(X_1= 2)=0.135$. The others
have marginal distributions $P(X_j=0)=(1-\pi_j)^2$,
$P(X_j=1)=2\pi_j(1-\pi_j)$ and $P(X_j=2)=\pi_j^2$, with $\pi_j$ ($j=3,
4, \ldots, 100$) being independently simulated from a beta
distribution with density $f(x) \propto x(1-x)^2$.
The models for $Y$ are:
\begin{eqnarray*}
  \mbox{M1:} \quad P(Y=1|\mathbf{X}) & = & 0.4 + 0.05I(Z=1) \{4I(X_1\neq 0) + 3I(X_2\neq 0)
  +I(X_1\neq 0, X_2\neq 0)\} \\
  \mbox{M2:} \quad P(Y=1|\mathbf{X}) &=& 0.3+0.2[\{2I(Z=1)-1\} I(X_1\neq 0, X_2\neq 0)\\
  && \mbox{} +I(X_3\neq 0)+I(X_4\neq 0)] \\
  \mbox{M3:} \quad P(Y=1|\mathbf{X}) & = & 0.5+0.1 [2\{I(Z=1)+I(X_1\neq 0)+I(X_2\neq 0)\} - 3].
\end{eqnarray*}
Figure~\ref{fig:models:m1&3} shows the values of $P(Y=1|\mathbf{X})$
for models~M1 and M3. Variables $X_1$ and $X_2$ are predictive in M1
but prognostic in M3.  Figure~\ref{fig:models:m2} shows the values for
model~M2 which is more complex; $X_1$ and $X_2$ are predictive and
$X_3$ and $X_4$ are prognostic. M2 tests the ability of a method to
distinguish between the two variable types.

\begin{figure}
  \centering
  \resizebox{0.8\textwidth}{!}{\rotatebox{-90}{\includegraphics*[44,37][585,751]{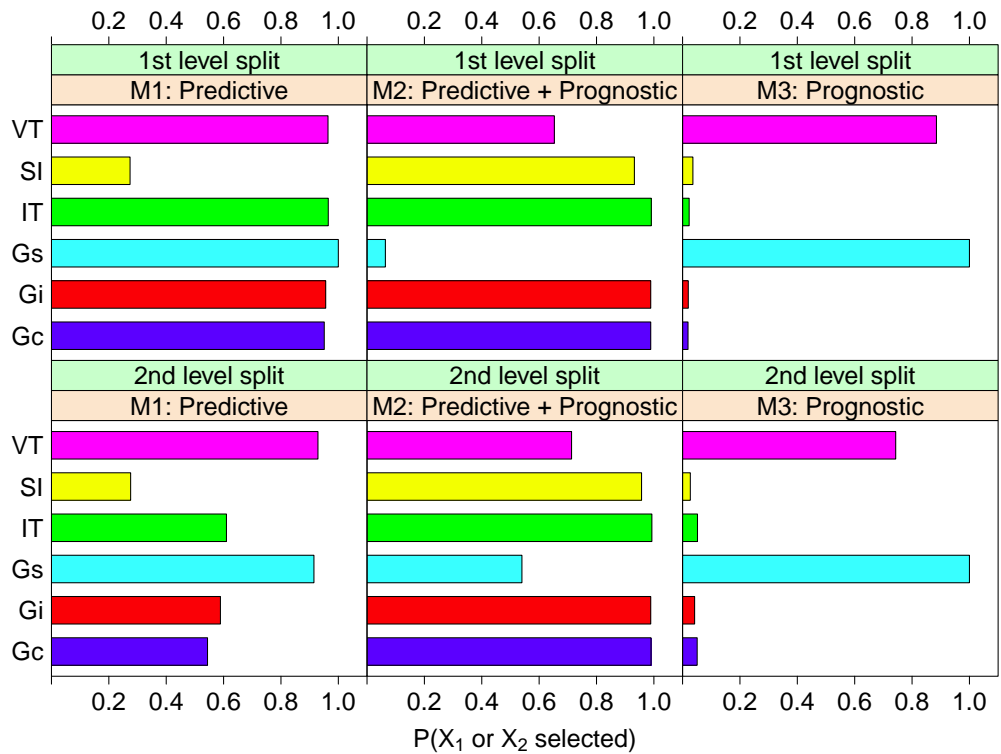}}}  
  \caption{Probabilities that $X_1$ or $X_2$ are selected at first and
    second level splits of trees for models M1, M2, and M3. Long bars
    are better for M1 and M2, and short bars are better for M3. SI
    refers to SIDES.}
  \label{fig:splits}
\end{figure}

First we compare the frequencies that $X_1$ and $X_2$ are chosen at
the first two levels of splits of a tree.  For each of 1000 simulation
iterations, 100 observations of the vector $(\mathbf{X}, Y, Z)$ are
simulated from each model and a tree is constructed using each
method. The frequencies that $X_1$ or $X_2$ is selected to split the
root node (1st level split) as well as one or both of its child nodes
(2nd level split) are graphed in Figure~\ref{fig:splits}.  QUINT is
excluded because it does not allow categorical variables and the
models do not have qualitative interactions. For model M1, where $X_1$
and $X_2$ are predictive and there is no other variable associated
with $Y$, all but SIDES select $X_1$ or $X_2$ to split the root node
with comparably high frequency. At the second split level, the
frequencies for Gs and VT are distinctly higher than those for Gc, Gi
and IT, while that for SIDES remains low. Therefore Gs and VT are best
and Gc, Gi and IT second best for model M1 on this criterion.  The
situation is different in M2 which has two predictive and two
prognostic variables. Now Gc, Gi, IT and SIDES are best and Gs is
worst and VT second worst. This shows that Gs and VT have difficulty
distinguishing between predictive and prognostic variables. This
behavior is repeated in M3 which has no predictive variables. Here the
probability that $X_1$ or $X_2$ is selected to split the nodes should
not be different from that of the other 98 variables, but Gs and VT
continue to pick the former with high frequencies. Only Gc, Gi, IT,
and SIDES perform correctly in this case.


Next we compare the power of the methods in identifying the correct
subgroup. Let $S$ be any subgroup. Recall from the Introduction that
the effect size is $R(S) = |P(Y=1 | Z=1, S) - P(Y=1 | Z=0, S)|$. The
``correct'' subgroup $S^*$ is defined as the maximal (in probability)
subgroup $S$ with the largest value of $R(S)$.  For models M1 and M2,
$S^* = \{X_1 \neq 0, X_2 \neq 0\}$; for M3, $S^*$ is trivially the
whole space because the effect size is 0.20 everywhere.

To estimate accuracy, let $n(t,y,z)$ denote the number of training
samples in node $t$ with $Y=y$ and $Z = z$ and define $n(t,+,z) =
\sum_y n(t,y,z)$ and $n_t = \sum_z n(t,+,z)$. Let $S_t$ be the
subgroup defined by $t$. The value of $R(S_t)$ is estimated by
$\hat{R}(S_t) = |n(t,1,1)/n(t,+,1) - n(t,1,0)/n(t,+,0)|$. The estimate
$\hat{S}$ of $S^*$ is the subgroup $S_t$ such that $\hat{R}(S_t)$ is
maximum among all terminal nodes.
If $\hat{S}$ is not unique, take their union.  The ``accuracy'' of
$\hat{S}$ is defined to be $P(\hat{S})/P(S^*)$ if $\hat{S} \subset
S^*$ and 0 otherwise.

\begin{figure}
  \centering
  \resizebox{0.8\textwidth}{!}{\rotatebox{-90}{\includegraphics*[42,37][581,752]{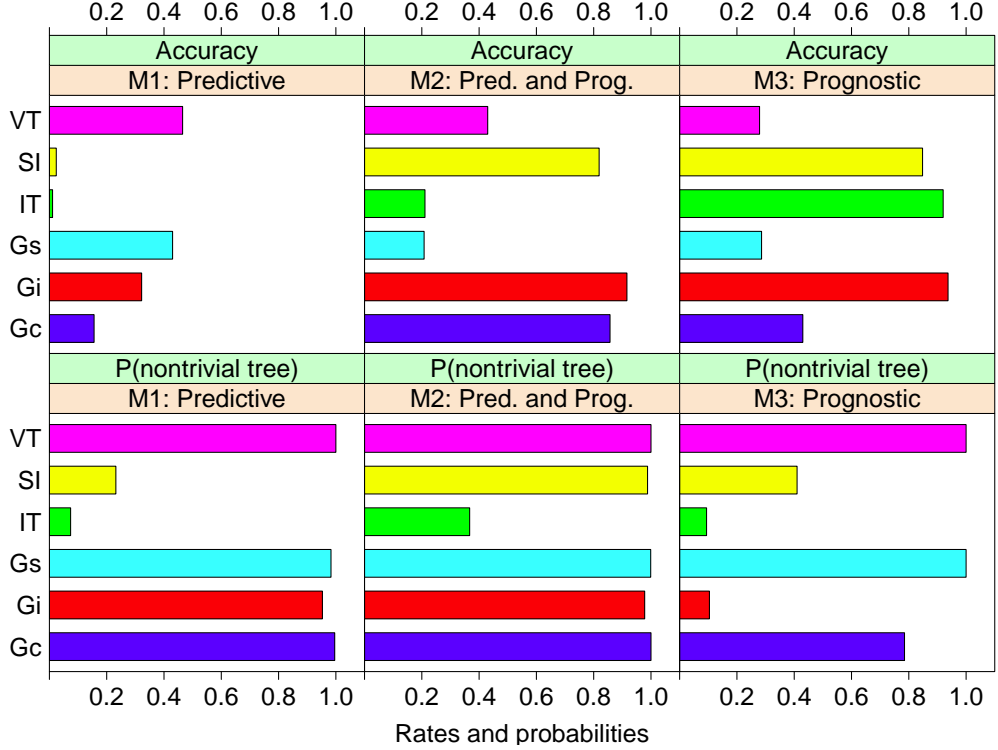}}} 
  \caption{Accuracy rates of subgroup identification and frequencies
    of nontrivial trees for models M1, M2, and M3. For accuracy, long
    bars are better.  For frequencies of nontrivial trees, long bars
    are better for M1 and M2 and short bars are better for M3. SI
    refers to SIDES.}
  \label{fig:power}
\end{figure}

\begin{table}
  \caption{Average computational times (sec.) for Model 1}
  \label{tab:speed}
  \centering
    \begin{tabular}{ccccccccc}
      \hline 
      MOB & Gs & Gi & Gc & IT & VT & SI & QU \\
      1.4 & 4.3 & 7.0 & 17.5 & 130.1 & 341.1 & 1601.5 & NA \\
      \hline 
    \end{tabular}
\end{table}

Table~\ref{tab:cv} and Figure~\ref{fig:power} show the estimated
accuracies and probabilities of nontrivial trees based on samples of
size 100 and 1000 simulation iterations. We see that:
\begin{description}
\item[Model M1.] Gs and VT are best, followed by Gi. Methods IT and
  SIDES have very low accuracy, due to their high tendency to yield
  trivial trees and hence no subgroups. The other four methods almost
  always give nontrivial subgroups.
\item[Model M2.] Gi has the highest accuracy, at 0.91. It is followed
  closely by Gc and SIDES at 0.86 and 0.82, respectively. Gs, IT and
  VT have difficulty distinguishing predictive from prognostic
  variables. All yield nontrivial trees almost all the time, except
  for IT which gives a trivial tree 63\% of the time.
\item[Model M3.] Because $S^*$ is the whole space, the ideal tree
  should be trivial always.  Methods Gi and IT are closest to ideal,
  yielding trivial trees 90\% of the time. In terms of accuracy, Gi,
  IT, and SIDES are best, having values of 0.94, 0.0.92, and 0.85,
  respectively. Gs and VT are the worst, because they produce
  nontrivial trees all the time.
\end{description}

The above results suggest that Gi is the overall best method in terms
of accuracy. It is best in models~M2 and M3 and third best in
model~M1, where it loses to Gs and VT, which are more accurate when
there are no prognostic variables.  Gc, IT and SIDES are dominated by
Gi in every model.

\section{Bootstrap confidence intervals}
\label{sec:boot}

Na\"{i}ve point and interval estimates of the treatment means and
differences can certainly be calculated from the training data in each
node.  Let $\mu(t,z)$ denote the true mean response for treatment $z$
in node $t$ and let $(y_i, z_i)$, $i =1, 2, \ldots, n_t$, be the
observations in $t$. Let $k_z$ denote the number of observations in
$t$ assigned to treatment $z$. Then $\hat{\mu}(t,z) =
k_z^{-1}\sum_{z_i=z} y_i$ is the na\"{i}ve estimate of $\mu(t,z)$; and if
$\hat{\sigma}(t,z)$ denotes the sample standard deviation of the $y_i$
among the treatment $z$ observations in $t$, then $\hat{\mu}(t,z) \pm
2 k_z^{-1/2}\hat{\sigma}(t,z)$ is a na\"{i}ve 95\% confidence interval for
$\mu(t,z)$.  Similarly, if $Z=0, 1$, let $d(t) = \mu(t,1) -
\mu(t,0)$. Then $\hat{d}(t) = \hat{\mu}(t,1)-\hat{\mu}(t,0)$ is the
na\"{i}ve estimate of the treatment effect and a na\"{i}ve confidence interval
for $d(t)$ is the usual two-sample t-interval.  Since the nodes in the
tree are not fixed in advance but are typically produced by a complex
optimization procedure, however, the validity of these estimates
should not be taken for granted.  For example, SIDES employs
adjustments to the na\"{i}ve p-values of treatment effects in the nodes to
control bias.

To see the extent of the bias for Gi and Gs, we carried out a
simulation experiment using models~M1 and M2. The experimental design
is an $r$-replicate ($r=2, 4$) of a $3^4$ factorial in variables $X_1,
X_2, X_3, X_4$, each taking values 0, 1, and 2, with $Z$ independent
Bernoulli with probability 0.50, and the binary response $Y$ simulated
according to models~M1 or M2.  In each simulation trial, a Gi or Gs
tree $T$ is constructed from the training data. If $T$ is nontrivial,
we record the average values of $\hat{\mu}(t,z)-\mu(t,z)$ and
$\hat{d}(t)-d(t)$ over the nodes $t$ and the proportions of times each
na\"{i}ve confidence interval contains the true estimand.  Columns~3--8 of
Table~\ref{tab:CI} show the estimated bias and coverage probabilities
of the intervals over 2000 simulation trials with nontrivial
trees. The biases are remarkably small (the true means range from 0.30
to 0.90). We attribute this to Gi and Gs being not directed at finding
splits to maximize or minimize the treatment effect, unlike SIDES and
QUINT. The coverage probabilities, on the other hand, are all too low,
although there is a perceptible improvement as $R$ increases..

We use the following method to construct better intervals by using the
bootstrap to estimate the standard deviations of the na\"{i}ve estimates.
Let $\mathcal{L}$ denote a given data set and let $T$ denote the
regression tree constructed from it.  Let $\mathcal{L}^*_j$
($j=1,2,\ldots, J$) be a bootstrap training sample from $\mathcal{L}$
and let $T^*_j$ be the tree constructed from $\mathcal{L}^*_j$ with
na\"{i}ve estimates $\hat{\mu}^*_j(t^*,z)$ for terminal nodes $t^*$ in
$T^*_j$.  Let $n_z(t \cap t^*)$ be the number of treatment $z$
observations from $\mathcal{L}$ that belong to $t \cap t^*$ and define
\[ \bar{\mu}^*_j(t,z) = \sum\nolimits_{t^*} n_z(t \cap t^*) \;
\hat{\mu}^*_j(t^*,z) \big/ \sum\nolimits_{t^*} n_z(t \cap t^*). \] The
bootstrap estimate of the variance of $\hat{\mu}(t,z)$ is the sample
variance $s^2_{\mu}(t,z)$ of $\{\bar{\mu}^*_1(t,z)$,
$\bar{\mu}^*_2(t,z)$, \ldots, $\bar{\mu}^*_J(t,z)\}$
and a 95\% bootstrap confidence interval for $\mu(t,z)$ is
$\hat{\mu}(t,z) \pm 2 s_{\mu}(t,z)$.  If $Z$ takes values 0 and 1, let
$\bar{d}^*_j(t) = \bar{\mu}^*_j(t,1)-\bar{\mu}^*_j(t,0)$.  Then a 95\%
confidence interval for $d(t)$ is $\hat{d}(t) \pm 2 s_d(t)$ where
$s^2_d(t)$ is the sample variance of $\{\bar{d}^*_1(t)$, $\bar{d}^*_2(t)$,
\ldots, $\bar{d}^*_J(t)\}$.

The rightmost three columns of Table~\ref{tab:CI} give the simulated
coverage probabilities of the bootstrap intervals using $J=100$.
There is a clear improvement over the na\"{i}ve intervals. In particular,
the bootstrap intervals for the treatment effect $d(t)$ are remarkably
accurate across the two models and two methods. The worst performance
occurs in model~M1 for $Z=0$, where the true treatment mean is 0.40 in
all nodes (see Figure~\ref{fig:models:m1&3}).

\section{Censored data}
\label{sec:censored} 
Several obstacles stand in the way of direct extension of Gi and Gs to
data with censored response variables.  The obvious approach of
replacing least squares fits with proportional hazards models in the
nodes \cite{NCASB05,SZYFY08} is problematic because Gi and Gs employ
chi-squared tests on residuals and their signs. Although there are
many definitions of such residuals \cite{ahnloh94}, it is unclear if
any will serve the purpose here. Besides, as noted earlier, fitting a
separate proportional hazards model in each node yields different
baseline cumulative hazard functions. As a result, the whole model no
longer has proportional hazards and hence regression coefficients
between nodes cannot be compared.  To preserve this property, a common
estimated baseline cumulative hazard function is required. We solve
these problems with the old trick of using Poisson regression to fit
proportional hazards models.

Let $u_i$ and $\mathbf{x}_i$ denote the survival time and covariate
vector of subject~$i$.  Let $s_i$ be an independent observation from
some censoring distribution and let $\delta_i = I(u_i < s_i)$ be the
event indicator.  The observed data vector corresponding to subject
$i$ is $(y_i, \delta_i, \mathbf{x}_i)$, where $y_i = \min(u_i, s_i)$,
$i=1,2,\ldots, n$.  Let $F(u, \mathbf{x})$ and $\lambda(u,
\mathbf{x})$ denote the distribution and hazard functions,
respectively, at $\mathbf{x}$. The proportional hazards model
specifies that $\lambda(u, \mathbf{x}) = \lambda_0(u) \exp(\eta)$,
where $\lambda_0(u)$ is the baseline hazard and $\eta =
\mathbf{\beta}'\mathbf{x}$ is a linear function of the covariates.
Let $\Lambda(u, \mathbf{x}) = \int_{-\infty}^u \lambda(z,
\mathbf{x})\,dz$ denote the cumulative hazard function and let
$\Lambda_0(u) = \Lambda(u, \mathbf{0})$ be the baseline cumulative
hazard. Then the density function is $f(u, \mathbf{x}) = \lambda_0(u)
\exp\{\eta - \Lambda_0(u)\exp(\eta)\}$. Letting $\mu_i =
\Lambda_0(y_i)\exp(\eta_i)$, the loglikelihood can be expressed as
\begin{eqnarray*}
  \lefteqn{\sum_{i=1}^n \delta_i \log f(y_i, \mathbf{x}_i) + 
    \sum_{i=1}^n (1-\delta_i) \log\{1-F(y_i, \mathbf{x}_i)\}} \\
  & = & \sum_{i=1}^n \left[\delta_i \{\log
    \Lambda_0(y_i)+\eta_i\}-\Lambda_0(y_i)\exp(\eta_i) + \delta_i
    \log\{\lambda_0(y_i)/\Lambda_0(y_i)\} \right] \\
  & = & \sum_{i=1}^n (\delta_i \log \mu_i - \mu_i) + \sum_{i=1}^n \delta_i 
  \log\{\lambda_0(y_i)/\Lambda_0(y_i)\}.
\end{eqnarray*}
The first term on the right is the kernel of the loglikelihood for $n$
independent Poisson variables $\delta_i$ with means $\mu_i$ and the
second term is independent of the covariates (see, e.g.,
\cite{aitkin80,Laird81}).  If the $\Lambda_0(y_i)$ values are known,
the vector $\mathbf{\beta}$ may be estimated by treating the event
indicators $\delta_i$ as independent Poisson variables distributed
with means $\Lambda_0(y_i)\exp(\beta' \mathbf{x}_i)$.

Thus we can construct a proportional hazards regression tree by
iteratively fitting a Poisson regression tree \cite{clly95,leh06},
using $\delta_i$ as Poisson responses, the treatment indicators as
predictor variables, and $\log \Lambda_0(y_i)$ as offset variable.  Gi
and Gs employ loglinear model goodness-of-fit tests
\cite[p.212]{agresti07} to the fitted values to obtain the split
variables and split points.  At the first iteration, $\Lambda_0(y_i)$
is estimated by the Nelson-Aalen \cite{aalen78,breslow72} method.
After each iteration, the estimated relative risks of the observations
from the tree model are used to update $\Lambda_0(y_i)$ for the next
iteration (see, e.g., \cite[p.~361]{Lawless82}).  The results reported
here are obtained with five iterations.

Figure~\ref{fig:breast:gsgi} gives the results of applying these
techniques to the breast cancer data. Gi and Gs each splits the data
once, at $\mathtt{progrec} \leq 21$ and $\mathtt{pnodes} \leq 3$,
respectively.  The corresponding Kaplan-Meier curves in the figure
show that \texttt{progrec} is predictive and \texttt{pnodes} is
prognostic.  The 95\% bootstrap confidence intervals of $\exp(\beta)$,
the relative risk of hormone therapy versus no therapy, are shown
beside the terminal nodes of the trees. They are constructed as for
uncensored response data, with the regression coefficient $\beta$
replacing the mean response. Specifically, let $\mathcal{L}$ and $T$
denote the training sample and the tree constructed from it. Let
$\mathcal{L}^*_j$ and $T^*_j$ denote the corresponding $j$th bootstrap
sample and tree, for $j = 1,2,\ldots,J$. Let $\hat{\beta}(t)$ and
$\hat{\beta}^*_j(t^*)$ denote the estimates of $\beta$ in nodes $t \in
T$ and $t^* \in T^*_j$ based on $\mathcal{L}$ and $\mathcal{L}^*_j$,
respectively, and let $n(A)$ denote the number of cases in
$\mathcal{L}$ that belong to any set $A$.  Define $\bar{\beta}^*_j(t)
= \sum\nolimits_{t^*} n(t \cap t^*) \; \hat{\beta}^*_j(t^*) \big/
\sum\nolimits_{t^*} n(t \cap t^*)$. The bootstrap estimate of the
variance of $\hat{\beta}(t)$ is the sample variance $s^2_{\beta}(t)$
of $\{\bar{\beta}^*_1(t)$, $\bar{\beta}^*_2(t)$, \ldots,
$\bar{\beta}^*_J(t)\}$ and a 95\% bootstrap confidence interval for
$\beta(t)$ is $\hat{\beta}(t) \pm 2 s_{\beta}(t)$.

\section{Importance scoring and thresholding}
\label{sec:imp}

When there are many variables, it may be useful or necessary to reduce
their number by some form of variable selection.  One way to
accomplish this is to rank them in their order of importance and
select a top-ranked subset.  Lack of a proper definition of
``importance'' has led to many scoring methods being proposed. Few
methods include thresholds for identifying the noise variables.  For
example, CART and random forest use the information from surrogate
splits to compute scores but not thresholds.

Following \cite{lchen}, we score the importance of a variable $X$ in
terms of the 1-df chi-squared statistics computed during variable
selection. Specifically, let $q_t(X)$ be the value of $q(X)$ (see
Algorithms~\ref{alg:Gs} and \ref{alg:Gi}) at node $t$ and $n_t$ be the
number of observations in $t$. We define the importance score of $X$
to be $\mbox{Imp}(X) = \sum_t n_t q_t(X)$ and approximate its null
distribution with a scaled chi-squared using the Satterthwaite method
\cite{satt46}.  This procedure is similar to that in \cite{lchen}
except for two differences. First, the latter employs the weight
$\sqrt{n_t}$ is used instead of $n_t$ in the definition of
$\mbox{Imp}(X)$.  The new definition increases the probability that
the variable selected to split the root node is top-ranked. The other
change is in the choice of threshold.  In \cite{lchen}, the threshold
is the $K^{-1}(K-1)$-quantile of the approximating distribution of
$\mbox{Imp}(X)$, where $K$ is the number of predictor variables, the
motivation being that $1/K$ of the unimportant variable are expected
to be found important. It is difficult to compute the
$K^{-1}(K-1)$-quantile of the distribution, however, if $K$ is
large. Therefore the threshold is defined to be the 0.95-quantile
instead. With this definition, Gi identifies only \texttt{progrec} as
important. The important variables for Gs are \texttt{pnodes},
\texttt{progrec}, and \texttt{estrec}, in descending order.


\section{Application to data with missing values}
\label{sec:gene}
Missing values pose two problems for tree construction.  The first is
how to deal with them during split selection and the second is how to
send observations with missing values through a split. CART uses a
system of surrogate splits that is biased toward choosing variables
with few missing values \cite{cruise,strobl07b}.
For variable selection, Gc, Gi and Gs create a ``missing'' category in
each contingency table.  Each split has the form $x \in S$, where the
set $S$ may contain the missing values.  There is some evidence that
this technique is best among classification tree methods if the
response variable takes two values \cite{ding10}.

We illustrate the method on a real data set from a retrospective
candidate gene study. Owing to confidentiality reasons, the data and
solutions are described in general terms here. A total of 1504
subjects were randomized to treatment or placebo and the response is
survival time in days, with sixty-three percent censored.  The
explanatory variables consist of 17 continuous-valued baseline
measures (\texttt{a1}, \texttt{a2}, and \texttt{b01}--\texttt{b15})
and 288 categorical variables, of which 6 are baseline measures
(\texttt{c0}--\texttt{c5}) and the rest are genetic variables
(\texttt{g001}--\texttt{g282}), each with two or three levels.  More
than 95\% (1435/1504) of the subjects have missing values in one or
more explanatory variables and only 7 variables (\texttt{a1, a2, b3,
  c0, c4, b15, g272}) are completely observed.

Although the treatment effect is statistically significant, its
magnitude is small.
The question is whether there are subgroups for which there are larger
treatment effects.  Owing to the large number of variables, a
traditional Cox proportional hazards model is inapplicable without
some sort of variable selection, even if restricted to the subset of
complete observations.

\begin{figure}
  \centering
  \psset{linecolor=black,tnsep=2pt,tnheight=0cm,treesep=3cm,levelsep=30pt,radius=10pt}
  \pstree[treemode=D]{\TC~[tnpos=l]{a2 $\leq$ 0.1 or NA}}{
    \TC[fillcolor=cyan,fillstyle=solid]~[tnpos=l]{863}
    ~{(0.73, 1.54)}
    \TC[fillcolor=cyan,fillstyle=solid]~[tnpos=r]{641}
    ~{(0.45, 0.81)}
  }
  \\ \vspace{1em}
  \resizebox{0.8\textwidth}{!}{\rotatebox{-90}{\includegraphics*[33,32][310,776]{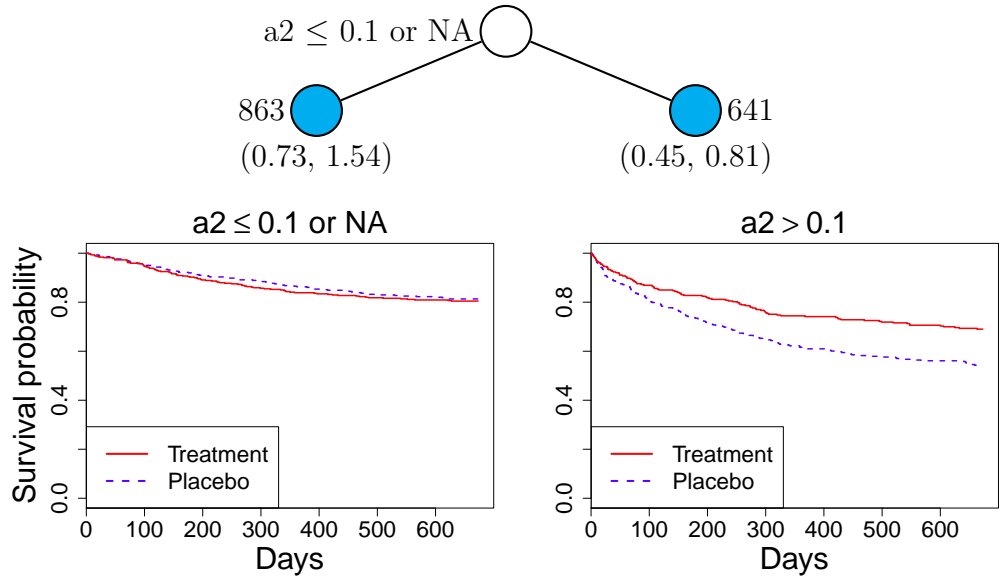}}}
  \caption{Gs model for gene data.  At each node, a case goes to the
    left child node if and only if the stated condition is satisfied.
    Sample sizes are beside terminal nodes and 95\% bootstrap
    intervals for relative risk of recurrence for treatment versus
    placebo are below the nodes.}
  \label{fig:gene:gs}
\end{figure}

The Gs model, shown in Figure~\ref{fig:gene:gs}, splits only once, on
\texttt{a2}. If the latter is less than 0.1 or missing, there is
little difference in survival probability between treated and
untreated, as shown by the Kaplan-Meier curves below the
node. Otherwise, the difference is statistically significant: a 95\%
bootstrap confidence interval (based on 100 bootstrap iterations) for
relative risk (treatment vs.\ placebo) is (0.45, 0.81).  The
importance scoring method identifies 27 and 28 important variables for
Gi and Gs, respectively.  The trees constructed from these variables,
however, are unchanged.


\section{Conclusion}
\label{sec:conc}



Regression trees are natural for subgroup identification because they
find subgroups that are interpretable.  But interpretability is
advantageous only of the algorithms that produce the trees do not
possess selection bias. We have introduced three algorithms that are
practically unbiased in this respect.  Gc is simplest because it can
be implemented with any classification tree algorithm (preferably one
without selection bias) by appropriate definition of a class
variable. It is limited, however, to binary-value response and
treatment variables. Further, some modification of the classification
tree algorithm may be needed to disallow splits that yield nodes with
pure classes.

Gs is a descendant of the GUIDE regression tree algorithm. As a
result, it is more general than Gc, being applicable to any kind of
ordinal response variables, including those subject to censoring, to
multi-valued treatment variables, and to all types of predictors
variables, with missing values or note. If there is no censoring, Gs
borrows all the ingredients of GUIDE. The main differences lie in the
use of the treatment variable as the sole predictor in a linear model
fitted to each node, the construction of a separate chi-squared test
of the residuals versus each predictor for each treatment level, and
the sum of the Wilson-Hilferty transformed chi-squared statistics to
form a single criterion for split variable selection at each node. As
the example in Figure~\ref{fig:regex:prog} demonstrates, however, Gs
can be distracted by the presence of prognostic variables.

Gi is our preferred solution if the goal is to find subgroups defined
by predictive variables only. To avoid being distracted by prognostic
variables, Gi uses a chi-squared test of treatment-covariate
interaction to select a split variable at each node. It is therefore
similar in spirit to the IT method. But unlike the latter, which
searches for the split variable and the split point at the same time,
Gi uses the chi-squared test for variable selection only. Besides
avoiding selection bias, this approach yields the additional benefit
of reduced computation time.

We extend Gi and Gs to censored time-to-event data by fitting a
tree-structured proportional hazards model to the data by means of
Poisson regression.  Poisson residuals are easier to use for our
purposes than those from proportional hazards models. Further, this
technique gives a common baseline cumulative hazard function and
allows comparisons of estimates between nodes.  The price is increased
computing time due to the need for iterative updates of the estimated
baseline cumulative hazard function, but the expense is not large
relative to the other methods, as shown by the average computing times
to construct one tree for model~M1 in Table~\ref{tab:cpu}. (Model~M1
has only $X$ variables that take three values each; the relative
speeds of Gc, Gi and Gs will be greater if these variables take more
values.)

Subgroup identification is prone to error if the number of predictor
variables greatly exceeds the sample size, because the chance of
finding the correct variables can be small, as our simulation results
demonstrate. If the number of variables is large, it is often helpful
to eliminate some of the irrelevant variables with importance score
thresholds and then construct the trees with the remaining ones.  Our
scoring and thresholding method is particularly convenient for this
purpose because, unlike other approaches, it does not require data
resampling and hence is much quicker. 

To our knowledge, there has not been an effective method of confidence
interval construction for the estimates in the nodes of a regression
tree. The main difficulty is the numerous levels of selection
typically involved in tree construction. Not surprisingly, na\"{i}ve
intervals that ignore the variability due to selection are overly
optimistic.  To solve this problem, we have to account for this extra
variability. We do this by using a bootstrap method to estimate the
true standard errors of the estimated treatment effects. Because each
bootstrapped tree is likely different (and different from the
original), we do not obtain an interval for each of its
nodes. Instead, we average the bootstrap treatment effects within each
node of the original tree and use the averages to estimate the
standard errors of the original treatment effects.  We do not yet have
theoretical proof of the consistency of this procedure, but the
empirical results are promising.

The Gi and Gs methods are implemented in the GUIDE computer program
which can be obtained from \url{www.stat.wisc.edu/~loh/guide.html}.

\section*{Acknowledgments}

We are very grateful to Xiaogang Su, Jared Foster, Jue Hou, and Elise
Dusseldorp for sharing with us their R programs for IT, VT, SIDES, and
QUINT, respectively, and for their patience in answering our numerous
questions.  We also thank Lei Shen for helpful comments on the
manuscript.  This work was partially supported by U.S. Army Research
Office grant W911NF-09-1-0205, NSF grant DMS-1305725, NIH grant
P50CA143188, and a grant from Eli Lilly and Company.





\bibliography{ref}

\clearpage

\begin{table}
  \caption{Four types of distributions of $X_1$ and $X_2$.}
  \label{tab:bias:d}
  \begin{center}
    \begin{tabular}{l|l|l} \hline
Notation & Type & Distribution   \\ \hline
Cont & Continuous  & Standard normal \\
Ord4 & Ordinal     & Discrete uniform with 4 levels \\
Cat3 & Categorical & Discrete uniform with 3 levels \\
Cat7 & Categorical & Discrete uniform with 7 levels \\
 \hline
\end{tabular}
\end{center}
\end{table}

\begin{table}
  \caption{Accuracy rates of subgroup selection and frequencies of nontrivial trees.
    Larger values are better. }
  \label{tab:cv}
\begin{center}
\begin{tabular}{clcccccc} \hline
Model& Type           &Gi    &Gs    &Gc    &IT    &SIDES&	VT\\
M1   & Accuracy  &0.322 &0.430 &0.150 &0.011 &0.024&	0.465\\
M1   & P(nontrivial tree)    &0.953 &0.983 &0.990 &0.074 &0.232&	1.000\\
M2   & Accuracy  &0.913 &0.204 &0.855 &0.211 &0.819&	0.430\\
M2   & P(nontrivial tree)    &0.979 &0.999 &1.000 &0.367 &0.988&	1.000\\
M3   & Accuracy  &0.939 &0.285 &0.519 &0.920 &0.848&	0.279\\
M3   & P(nontrivial tree)    &0.104 &1.000 &0.693 &0.094 &0.410&	1.000\\
 \hline
\end{tabular}
\end{center}
\end{table}

\begin{table}
  \caption{Bias of estimated treatment means and their difference (averaged over
    nodes of each tree) and coverage probabilities of na\"{i}ve and
    bootstrap 95\% intervals, based on 1000 simulations trials of nontrivial trees,
    with 100 bootstrap iterations per trial. $r$ is the number of replicates of 
    a $3^4$ design.}
  \label{tab:CI} \vspace{0.5em}
  \centering
    \begin{tabular}{cc|rrc|ccc|ccc}
      & & \multicolumn{3}{c|}{Bias of na\"{i}ve estimates} 
      & \multicolumn{6}{c}{Coverage probabilities of 95\%} \\
      & & \multicolumn{3}{c|}{of means and difference} 
      & \multicolumn{3}{c|}{na\"{i}ve intervals} 
      & \multicolumn{3}{c}{bootstrap intervals} \\
      $r$ & Expt & $\mu(t,0)$ & $\mu(t,1)$ & $d(t)$  
      & $\mu(t,0)$ & $\mu(t,1)$ & $d(t)$  
      & $\mu(t,0)$ & $\mu(t,1)$ & $d(t)$ \\ \hline
      2 & M1-Gi & 6.8E-3 & -2.2E-2 & -2.9E-2 & 0.821 & 0.811 & 0.818 & 0.892 
      & 0.955 & 0.934 \\
      & M1-Gs & 4.4E-3 & -1.9E-2 & -2.3E-2 & 0.819 & 0.800 & 0.857 
      & 0.907 & 0.952 & 0.935 \\
      & M2-Gi & 6.8E-3 & -2.0E-2 & -2.6E-2 & 0.835 & 0.846 & 0.836 & 0.937 &
      0.947 & 0.941 \\
      & M2-Gs & 3.5E-4 & -1.5E-2 & -1.5E-2 & 0.871 & 0.861 & 0.907 & 0.953 & 0.965
      & 0.942 \\ \hline
      4 & M1-Gi & 3.0E-2 & -1.6E-2 & -1.9E-2 & 0.880 & 0.874 & 0.889
      & 0.903 & 0.972 & 0.957 \\
      & M1-Gs & 3.7E-3 & -1.5E-2 & -1.9E-2 & 0.869 & 0.862 & 0.888 & 0.916 & 0.967
      & 0.955 \\
      & M2-Gi & 1.1E-3 & -7.5E-3 & -8.6E-3 & 0.896 & 0.915 & 0.911
      & 0.966 & 0.967 & 0.963 \\
      & M2-Gs & -3.8E-3 & -9.4E-3 & -5.7E-3 & 0.888 & 0.913 & 0.916
      & 0.968 & 0.973 & 0.950 \\
      \hline
    \end{tabular}
\end{table}

\begin{table}
  \caption{Average times (sec.), over 500 simulation trials  of model~M1, to 
    construct one tree on a 2.66GHz Intel processor; QUINT does not allow
    categorical variables} \label{tab:cpu}
  \begin{center}
    \begin{tabular}{cccccccc}
      \hline 
      Gs & Gi & Gc & IT & VT & SIDES & QUINT \\
      4.3 & 7.0 & 17.5 & 130.1 & 341.1 & 1601.5 & NA \\
      \hline 
    \end{tabular}
  \end{center}
\end{table}

\end{document}